\documentclass[preprint,5p,times]{elsarticle}
\usepackage[T1]{fontenc}
\usepackage[utf8]{inputenc}
\usepackage{graphicx}
\usepackage{textcomp}
\usepackage{siunitx}
\usepackage{tabularx}

\journal{NIM A}

\begin{document}
\begin{frontmatter}
\title{Tailoring phase-space in neutron beam extraction}
\author[frm]{S. Weichselbaumer}
\author[frm,e21]{G. Brandl}
\author[frm,e21]{R. Georgii}
\ead{Robert.Georgii@frm2.tum.de}
\author[lns]{J. Stahn}
\author[ldm]{T. Panzner}
\author[e21]{P. B\"{o}ni}
\address[frm]{Heinz Maier-Leibnitz Zentrum und Physik-Department E21,
  Technische Universit\"{a}t M\"{u}nchen, Lichtenbergstr.\ 1, D-85748 Garching, Germany}
\address[e21]{Physik-Department E21, Technische Universit\"{a}t M\"{u}nchen,
  James-Franck-Str.\ 1, D-85748 Garching, Germany}
\address[lns]{Laboratory for Neutron Scattering, Paul Scherrer Institut, CH-5232 Villigen PSI, Switzerland}
\address[ldm]{Material Science and Simulations, Neutrons and Muons, Paul Scherrer Institut, CH-5232 Villigen PSI, Switzerland}

\date{\today}

\begin{abstract}
In view of the trend towards smaller samples and experiments under extreme
conditions it is important to deliver small and homogeneous neutron beams to the
sample area. For this purpose, elliptic and/or Montel mirrors are ideally suited as the phase
space of the neutrons can be defined far away from the sample. Therefore, only
the useful neutrons will arrive at the sample position leading to a very low
background. We demonstrate the ease of designing neutron transport systems using
simple numeric tools, which are verified using Monte-Carlo simulations that
allow to take into account effects of gravity and finite beam size. It is
shown that a significant part of the brilliance can be transferred from the
moderator to the sample. Our results may have a serious impact on the design of
instruments at spallation sources such as the European Spallation Source (ESS)
in Lund, Sweden.
\end{abstract}
\end{frontmatter}

\section{Introduction}
The foundation laying  for the European Spallation Source (ESS) in Lund, Sweden
took place in October 2014. ESS is intended to operate at a power of 5 MW and
will use a long-pulse target station for the neutron production. The resulting,
time-integrated flux will be comparable or even larger than the continuous flux at the high flux
reactor HFR at the Institut Laue-Langevin in Grenoble \cite{yellowbook1986}. However, the
peak flux at ESS will exceed the time-averaged flux of the ILL by at least
a factor of 30. Therefore, using time-of-flight techniques the performance will
be largely increased. Further increases will be possible by implementing modern
neutron transport systems based on non-linearly tapered neutron guides and a
clever design of the instruments.

For more than three decades, with the invention of neutron guides by
Maier-Leibnitz \cite{MaierLeibnitz1963217}, neutrons were transported over large
distances mostly by Ni-coated, straight or curved guide tubes. However, due to
the small critical angle of total reflection given by $\theta\, /\, ^\circ =
0.099m\lambda$ / \AA, where $m = 1$ for Ni, the transport was only efficient for
cold neutrons. Using supermirror coatings, the index $m$ was increased up to $m
= 7$ \cite{swissneutronics} thus allowing to even transport epithermal neutrons with wavelength of $1\AA$
at spallation sources. 

Due to the many internal reflections of the neutrons in straight high-m neutron guides,
however, the transmission is seriously reduced \cite{boni2010}. Moreover, the
significant losses require massive shielding of the neutron guides. Mezei \cite{Mezei1997} and 
Schanzer et al.\  \cite{Schanzer200463} proposed the use of a truly bent elliptic
neutron guides, which reduce the number of reflections to essentially two. These guides 
are focusing the neutrons from the moderator exit to the sample  in terms of the
point to point imaging provided by an ellipse in mathematics \cite{cussen2013}.
For example, the replacement of the straight neutron guide
at the beam line HRPD at ISIS by a 90 m long elliptic guide increased the
neutron flux at the sample by up to two orders of magnitude
\cite{ibberson2009}.  In addition, as the beam paths can be simply identified
using geometrical optics, it is straightforward to design and judge the
performance of elliptic guides \cite{boeni2014}.  Recently Klen\o\ et al.\ have
shown that approximately 50\% to 90\% of the brilliance of cold (4.25 \AA~ $\le
\lambda \le$ 5.75 \AA) and thermal (0.75 \AA$\le \lambda  \le$ 2.25 \AA) neutrons
can be transported from the moderator to the sample using
parabolic or elliptic guide geometries \cite{Kleno2012}.  Effects of gravity
were included in this study.

Often, it is argued that elliptic guides are prone to a large background at the
sample because there is a direct view to the moderator. However, by inserting
beam blockers in the central part of the guide, the line of sight can be
effectively interrupted \cite{boni2008} without affecting the homogeneity of the
beam. It is correct, that the blocker leads to a hole in the transmitted phase
space as pointed out by Zendler et al.\ \cite{zendler2014}. However, this hole is
very small, as for instance in our simulation of the order  of \ang{0.12} (see
Fig.\ref{psd}b) and is therefore swamped by effects of waviness if
real neutron
guide systems are considered \cite{boeni2014}.  The major background of
elliptic guides is caused by the fast neutrons that emerge from the neutron
source and illuminate the internal surfaces of the guide close to the sample
\cite{boeni2014}.  These neutrons can effectively be stopped by placing two or
more elliptic guides in series \cite{boni2008} with the further advantage that
effects of halo and coma aberration are reduced, if an even number of guides is
used. Then even beam blockers may become superfluous. The direct line of sight
can also be interrupted by gravitational curving of long neutron guides
\cite{kleno2011}.

Amongst the other guide concepts, Montel guides are very promising in
    delivering neutrons to the sample. These mirrors were invented by Montel in 1957 for focusing X-rays
\cite{Montel:1957} and have now become an integral part of many beam lines at
synchrotron sources and for x-ray diffractometers. Recently, Montel mirrors have
been used by G.~Ice for the focusing of neutron beams \cite{Ice:he5452}. Stahn
et al.\ use them for reflectometry, i.e. for the SELENE project \cite{stahn2012}.
In addition, a guide system based on Montel mirrors was optimized for a proposed
MIEZE type spin echo spectrometer for the ESS \cite{Weber201371}.

    A Montel mirror consists of two elliptic mirrors that are arranged perpendicular to each other, i.e. the optics
    consists of a ''quarter'' of a four-sided elliptic neutron guide. In the
    center of the Montel mirror a beamstop can be placed for defining the
    accepted divergence independent from the beam size which is defined by the entrance aperture. Because the Montel mirror is inclined in the horizontal
    and vertical direction the direct line of sight is interrupted leading to an
    excellent signal-to-noise ratio on the sample.

Montel mirrors have many advantages when compared with other concepts for neutron
guides. Beside of the advantage of tailoring the neutron beam more than
typically ten meters away from the sample position \cite{stahn2011}, the path of
the neutrons through the optical system is clear, i.e. it takes place
via two reflections in each device. Moreover, the brilliance transfer can
easily be evaluated based on reflectivity data of the supermirrors. 

\begin{figure}
  \begin{center}
    \includegraphics[width=\linewidth]{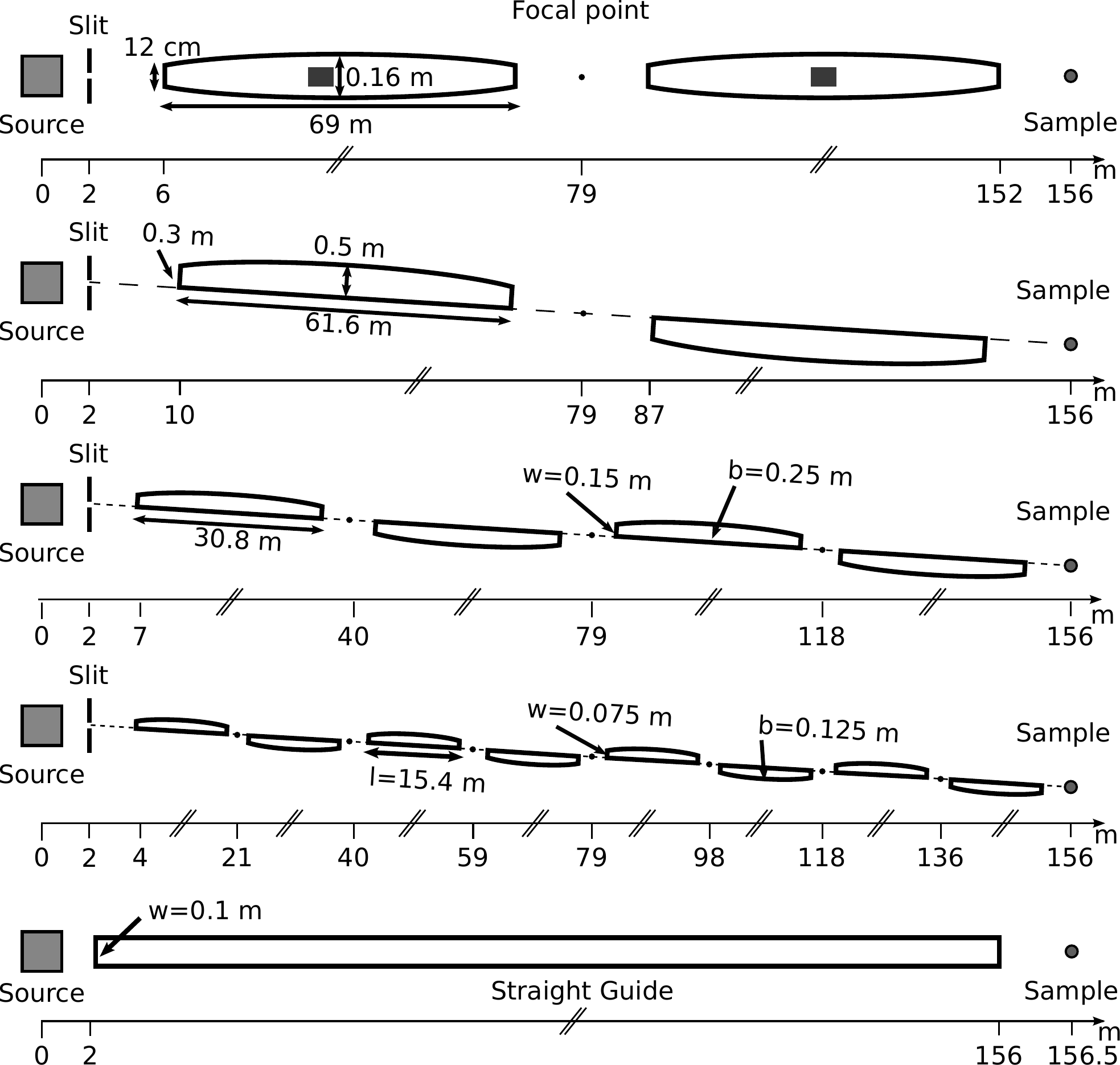}
     \caption{\label{NL}The different extraction methods and guide designs in this
     paper are shown in a top view for comparison. In order from top to bottom these
     are a double elliptic guide, the 2-fold and the 4-fold  Montel guides and a
     classical straight guide to get out of the direct sight of the moderator. All
     mirrors have a $m = 7$ coating. The beam stop in the elliptic guide has a
     height of \SI{0.14}{m} and is absorbing all neutrons, which have not been
     reflected. The Montel guides are inclined by \ang{1.25} in both horizontal
     and vertical direction (not drawn here).
     } 
  \end{center}
\end{figure}

The aim of the present work is, firstly, to evaluate the performance of various
types of neutron guides, including elliptic, Montel and straight guide systems,
using geometrical optics and analytical tools (Fig.~\ref{NL}). In a second step
we will verify the numeric results using the Monte-Carlo simulation package
McStas \cite{mcstashomepage}.  The results show that it is indeed possible to
calculate the performance for small beam and sample sizes and ignoring gravity rather accurately using
simple analytical means. Finally, gravitational effects will be taken into account using
Monte-Carlo simulations. 

\section{Numeric calculations without gravity}
\begin{figure}
  \begin{center}
    \includegraphics[width=\linewidth]{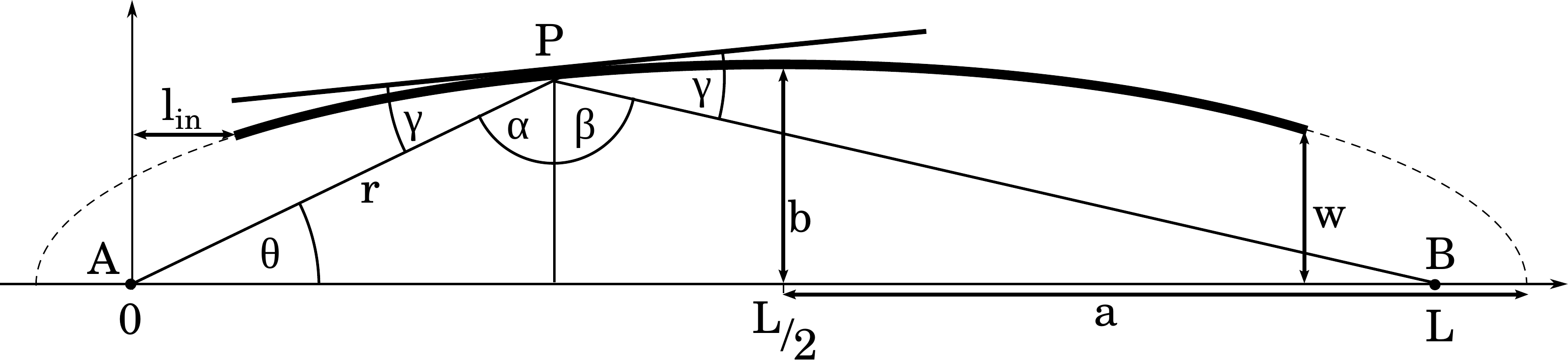}
 \caption{\label{fig:montel-schematic}Schematic view of an elliptic mirror. $a$
 and $b$ are the semi-major and semi-minor axis, respectively. $L$ is the
 distance between the focal points $A$ and $B$. The angles $\alpha$ and $\beta$
 can be used to calculate the reflection angle $\gamma$ of the neutron
 trajectory with the surface of the ellipse at the point $P$. $w$ and
$l_{\text{in}}$ are the distance of the guide to the central axis and the
distance between the focal point to the guide entry, respectively. }
  \end{center}
\end{figure}

In a first step we calculate the angle of reflection of neutrons emerging from a
point source at the first focal point $A$ of the ellipse, hitting the mirror at
the point $P$ and arriving at the second focal point $B$
(Fig.~\ref{fig:montel-schematic}). If the contour of the ellipse is represented
by a parametric equation in polar coordinates, the distance $r$ is given by 
\begin{equation}
    \overline{AP} = r(\theta) = \frac{a(1-e^2)}{1-e\cos\theta}
\end{equation}
where $e = L/2a$ is the numerical eccentricity of the ellipse, $L =
2\sqrt{a^2-b^2}$ is the distance between the focal points $A$ and $B$, and $a$
and $b$ are the half axes of the ellipse. The local angle of reflection,
$\gamma$, is given by
\begin{equation}
    \gamma =\frac{\pi}{2} - \frac{\alpha+\beta}{2}, 
\end{equation}
where
\begin{equation}
       \alpha = \frac{\pi}{2} - \theta,\quad \beta =
    \arccos\left({\frac{r\,\sin\theta}{2a-r}}\right).
\end{equation}
$\gamma$ can then be used to calculate the reflectivity $R'(\lambda, \gamma)$
of the supermirror in dependence of the neutron wavelength $\lambda$ and
eventually the transmitted intensity can be determined for all wavelengths.

In the following, a point-like source and a $m = 7$ supermirror (which is
currently state-of-the-art) is
assumed for all guides. The
reflectivity profile of the supermirror is parametrized by assuming a constant
reflectivity $R = 1$ up to a critical value $Q_c = 0.0219\,$\AA$^{-1}$
(corresponding to $m = 1$) followed by a linear decrease to
$R = 0.5$ at $q = m \cdot Q_c = 7 Q_c$. The transmitted neutron intensity for
a specific wavelength can be determined by integrating the reflectivity along
the whole mirror:
\begin{equation}
    R(\lambda) = \frac{1}{\theta_1 - \theta_0}\int_{\theta_0}^{\theta_1}
    R'(\lambda, \theta) \,\mathrm{d}\theta,
\end{equation}
where $\theta_0 = \arctan{(w/(L - l_{\text{in}}))}$ and $\theta_1 =
\arctan{(w/l_{\text{in}})}$ (see fig.~\ref{fig:montel-schematic}) are the angles
which define the accepted divergence.
$w$ and
$l_{\text{in}}$ are the distance of the guide to the central axis and the
distance between the focal point to the guide entry, respectively. To obtain
the intensity for $n$ Montel mirrors, the reflectivity has to be taken to the
power of $2n$ since a neutron is reflected twice by each Montel mirror.

The numeric results are valid in the limit of beams with no divergence. Therefore we verified them by using Monte Carlo simulations choosing
for the in and outgoing beam a size $5\times\SI{5}{mm^2}$ with a small divergence of $\pm 0.25^o$.
To compare the different guide concepts we use the brilliance transfer (BT) as it was defined
by Klen\o\ et al.\ \cite{Kleno2012}. The brilliance or phase space density
$\Psi$ is the number of neutrons per unit time, area, solid angle and
wavelength interval. According to Liouville's theorem the brilliance transfer
$\text{BT} = \Psi_\text{entry}/\Psi_\text{sample}$ can never be larger than
one. To measure the BT we place monitors with the same restriction in size, wavelength and
divergence after the entry slit (for the straight guide in front of the guide entry)
and at the sample position.

\begin{figure}
    \includegraphics[width=\linewidth]{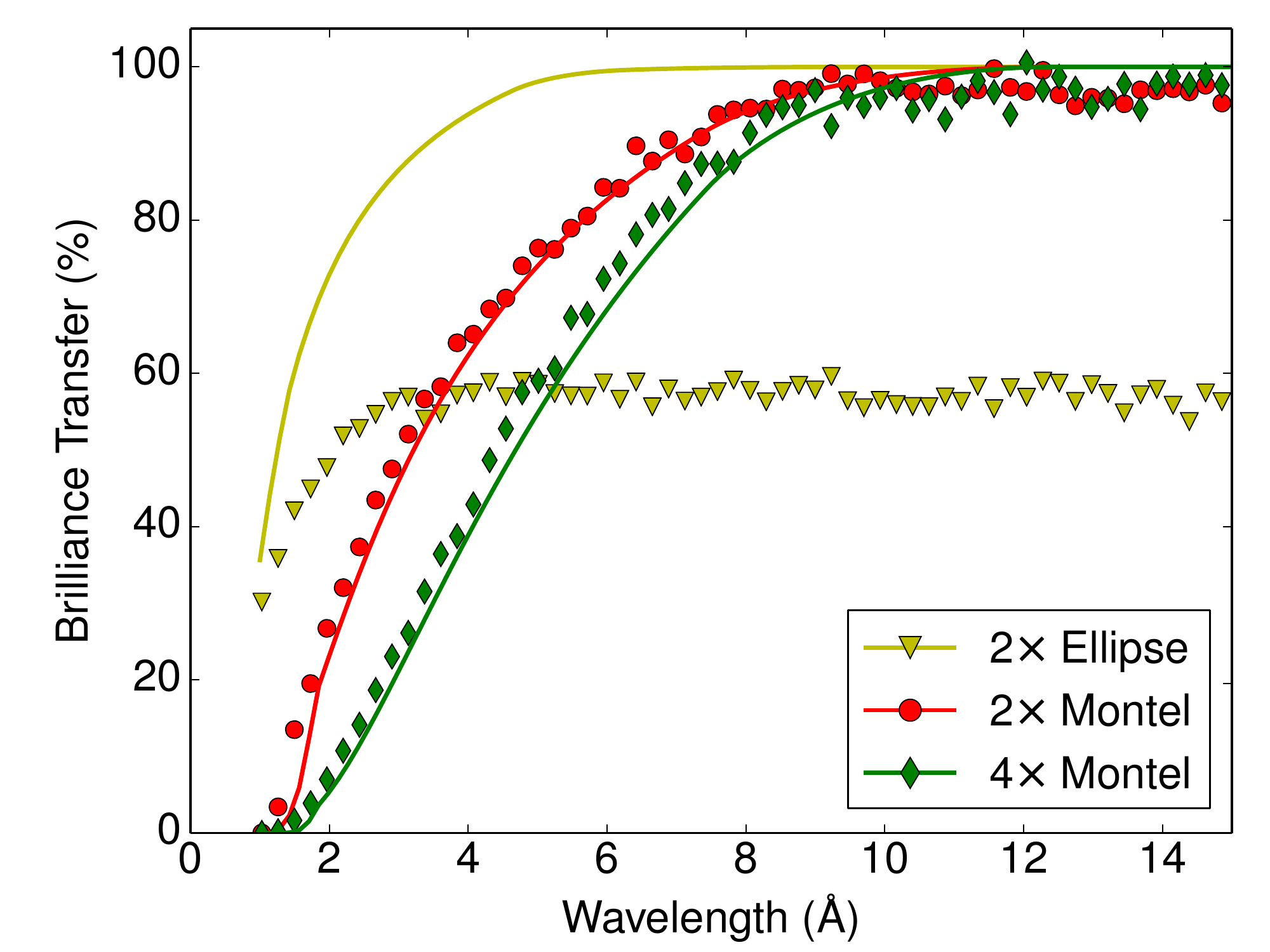}
    \caption{\label{bt}Comparison of the BT of the double elliptic guide
    system and the 2-fold and 4-fold Montel mirror systems (all $m = 7$) versus wavelength.  The solid lines are the
    analytically calculated curves and the data points are the result of 
    Monte-Carlo simulations. The simulations where performed without taking
    effects of gravity into account. The opening of the entry slit and
    the sample size is $5 \times\SI{5}{mm^2}$ and the divergence $\pm 0.25^o$. } 
\end{figure}

Fig.~\ref{bt} shows that the Monte Carlo simulations match the functional form of the numeric
predictions very nicely and predict in the case of the Montel optics even
correctly the BT. The smaller then predicted BT for the elliptic guides can be
attributed to the absorption of neutrons in the beam stop in the elliptic guides
which is not included in the analytical model. Due to the small divergence used this effect is 
particularly strong in Fig.~\ref{bt} and will be less prominent in the simulations with larger divergence of $\pm 1^o$ which will be used in the following section.


In the light of these results, whenever possible, numeric calculations for the design of neutron guides
should be conducted first to provide independent tests of the correct placement of the guides in the simulations. It allows also for first
estimations of the efficiency of a neutron guide system facilitating and speeding up the comparison of different designs. 
However, to include the effects of larger beam and  sample sizes and the effect of gravity  Monte-Carlo
simulations are required.

 \section{Inclusion of effects of gravity}
In the following we investigate the effect of gravity on the performance of
the transport systems discussed above. Gravitational effects are large: For
example, neutrons with $\lambda = 5$~\AA\ drop 193 mm along a free flight path
of 156 m. Due to the complexity of the problem, Monte-Carlo simulations are
mandatory.  

\begin{figure} \centering
    \includegraphics[width=0.9\linewidth]{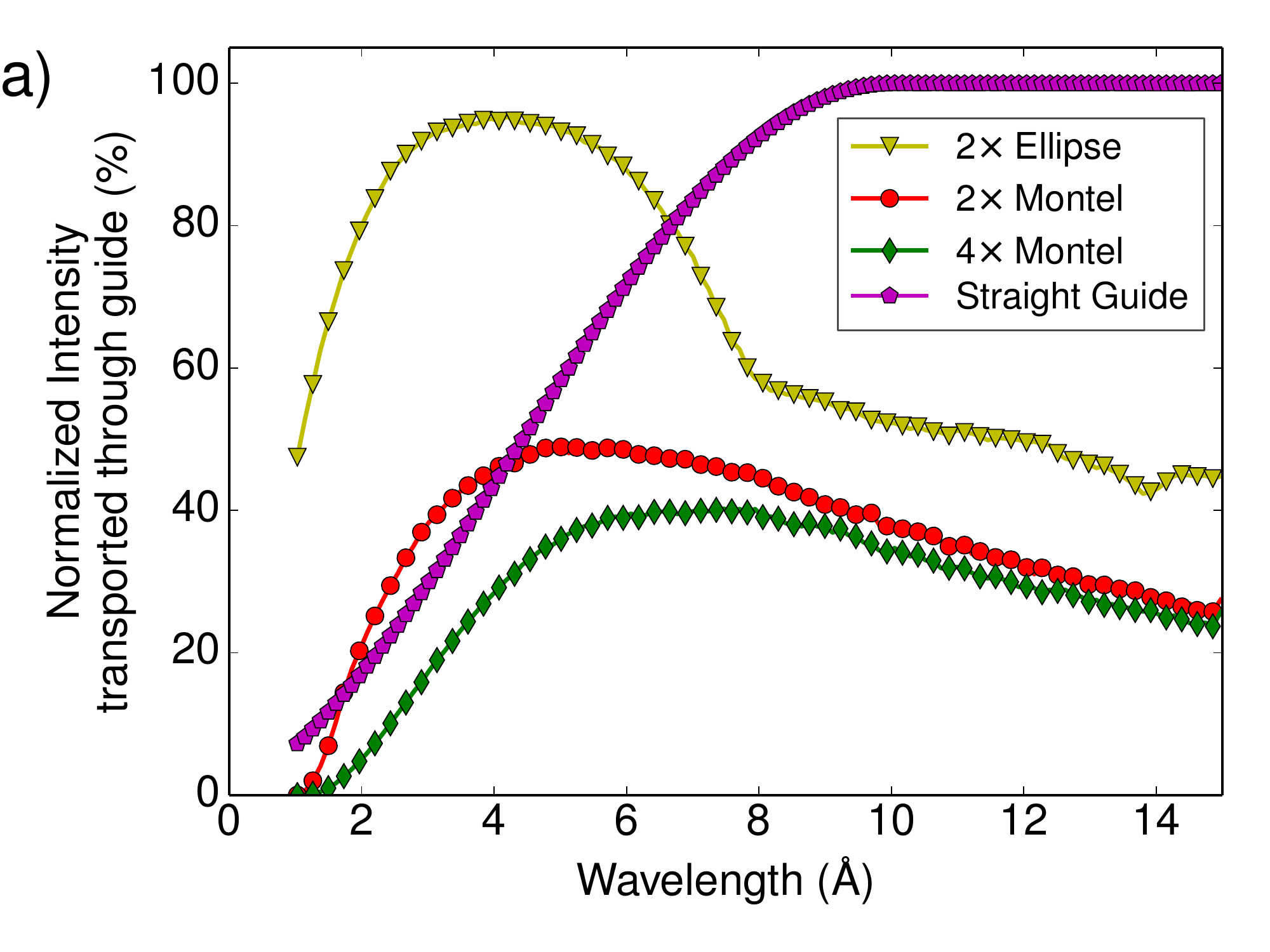}
    \includegraphics[width=0.9\linewidth]{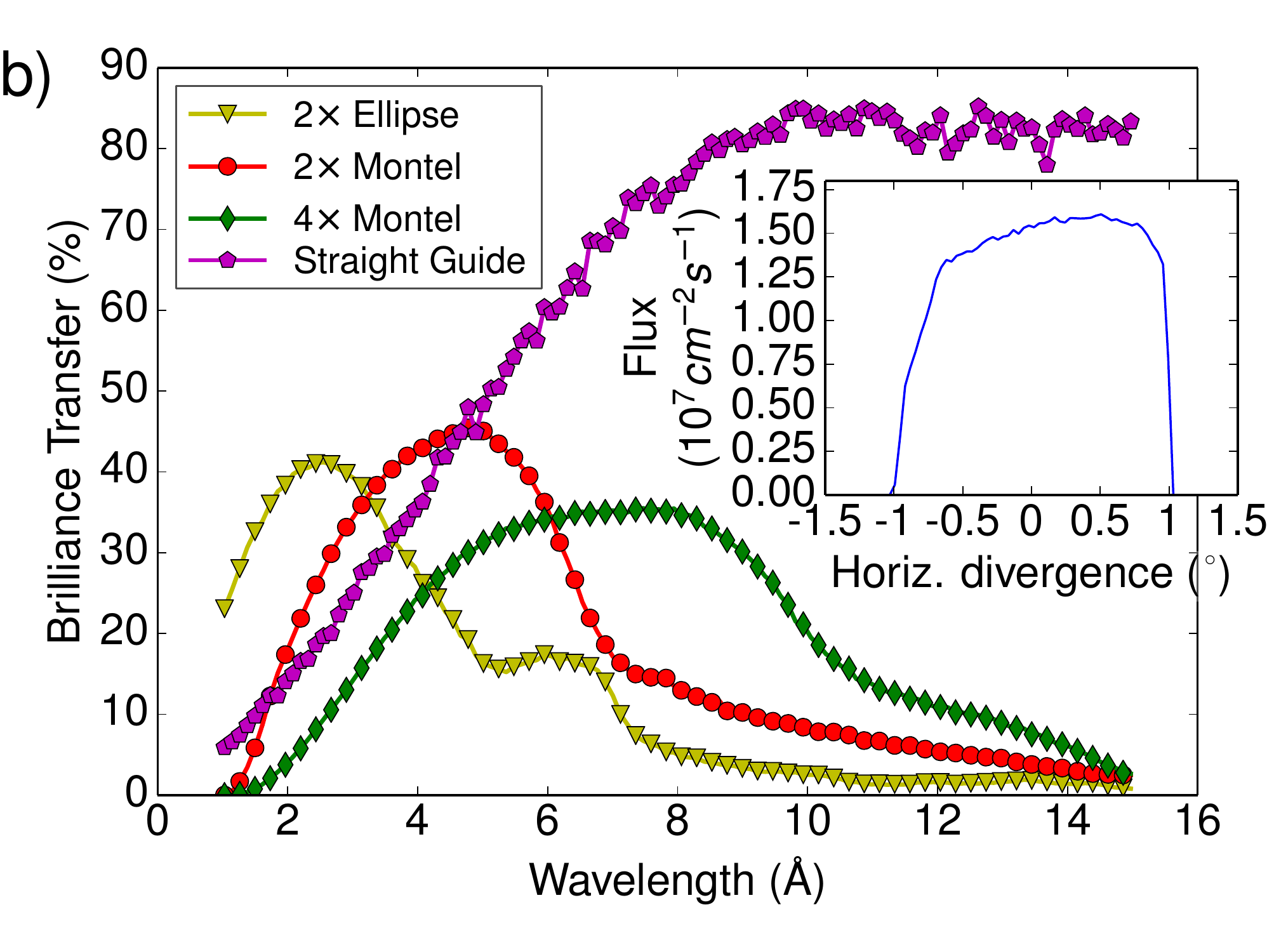} \caption{
        \label{fig:bt2x2}(a) The integrated intensity after the guide
        exit vs. wavelength normalized to the intensity at the guide entry and
        (b) the brilliance transfer vs.  wavelength  for different guide
        geometries respecting the effects of gravity. The inset
        shows the divergence distribution at the sample position. The BT for the
        Montel optics peaks at different $\lambda$ depending on the number of
        mirrors. The beam size and sample size was $5\times\SI{5}{mm^2}$. The
        inset shows the horizontal divergence of the 2-fold Montel system at the
        sample position. }
\end{figure}

For  all simulations in this section a flat wavelength spectrum with a brilliance of $\Psi = 1
\times 10^{12} $cm$^{-2} $s$^{-1} \AA^{-1}$ sterad$^{-1}$ for $\SI{1}{\AA} \le
\lambda \le \SI{15}{\AA}$ is considered. The divergence at the sample is assumed
to be $\pm 1^\circ$. An aperture with an opening corresponding to the assumed sample
size is placed at the position of the closest place for a neutron guide at the
ESS, i.~e.~$2$ meters away from the moderator (see Fig.~\ref{NL}).  For the
straight guide a cross section of ${100} \times \SI{100}{mm^2}$ and no entry
slit is assumed.  A quantitative comparison of the normalized intensity
transported through the guide system and the brilliance transfer (BT) for the
different neutron guide systems is shown in Fig.~\ref{fig:bt2x2} for an
aperture of $5\times\SI{5}{mm^2}$.  The difference here is, that the upper plot
shows the intensity normalized to the guide entry whereas the BT shows the
brilliance at the sample position normalized to the brilliance at the entry
slit.
In Tab.~\ref{tab:intensities} the effect of
gravity is given for the different guide systems as the ratio of the integrated
intensities without and with gravity. Furthermore the signal--to--noise ratio
(SNR), defined as ratio of the integrated intensity on a sample of size of
$5\times\SI{5}{mm^2}$ to the total transported integrated intensity is given.

\begin{table}
    \centering
    \small
    \begin{tabularx}{0.8 \columnwidth}{l c c c c}
       \hline
        \textbf{Config.} & \textbf{No gravity} & \textbf{Gravity} &
        \textbf{Ratio} & \textbf{SNR} \\
        & $I_{\text{NG}}$ ($\si{s^{-1}}) $
        & $I_{\text{G}}$ ($\si{s^{-1}}) $
        & $I_{\text{NG}}/I_{\text{G}}$ & ($\%$)\\
       \hline
        2 x Ell. & $5.7\times 10^8 $& $3.4 \times 10^8$ & $1.7$ & $57$ \\
        2 x Mon. & $6.\times 10^8$ & $4.6\times 10^8$ & $1.3$ & $90$ \\
        4 x Mon.& $3.3\times 10^8$ & $2.9\times 10^8$ & $1.1$ & $91$ \\
        Straight& $4.6\times 10^8$& $4.6\times 10^8$ & $1.0$ & $0.25$ \\
       \hline
    \end{tabularx}
     \caption{Integrated intensities on a $5\times\SI{5}{mm^2}$ sample for a
    wavelength band of \SI{2}{\AA}~--~\SI{6}{\AA} for the different configurations.
    A ratio close to $1$ means that gravity has only a small effect on the
    transport properties. The SNR (Signal-to-Noise ratio) specifies how much
    intensity arrives in the case for enabled gravity inside the sample
    rectangle shown in Fig.~{\ref{psd}}a compared to the overall intensity on the position sensitive
    detector (PSD).}
    \label{tab:intensities}
\end{table}

\begin{figure}
\centering
    \includegraphics[width=0.9\linewidth]{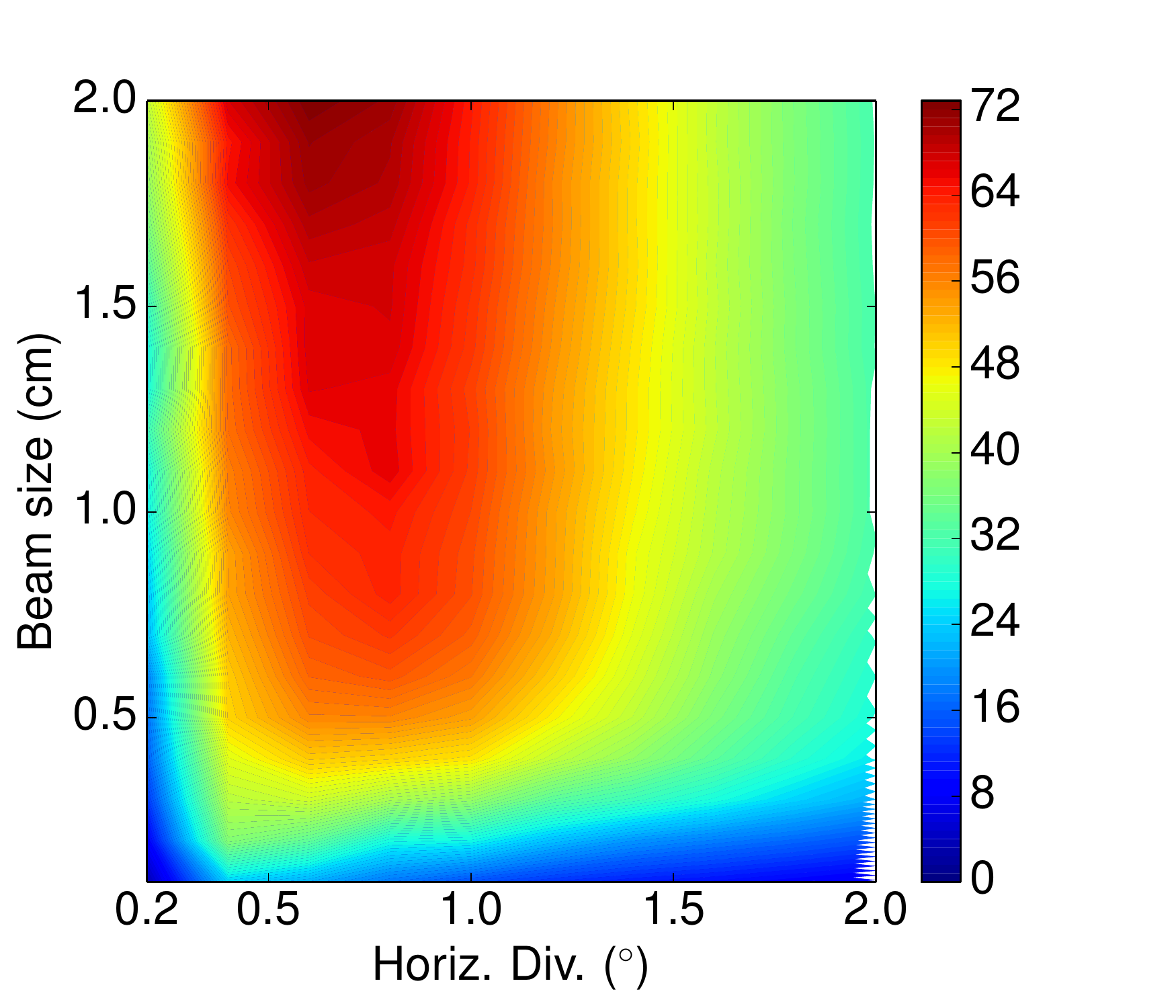}
    \includegraphics[width=0.9\linewidth]{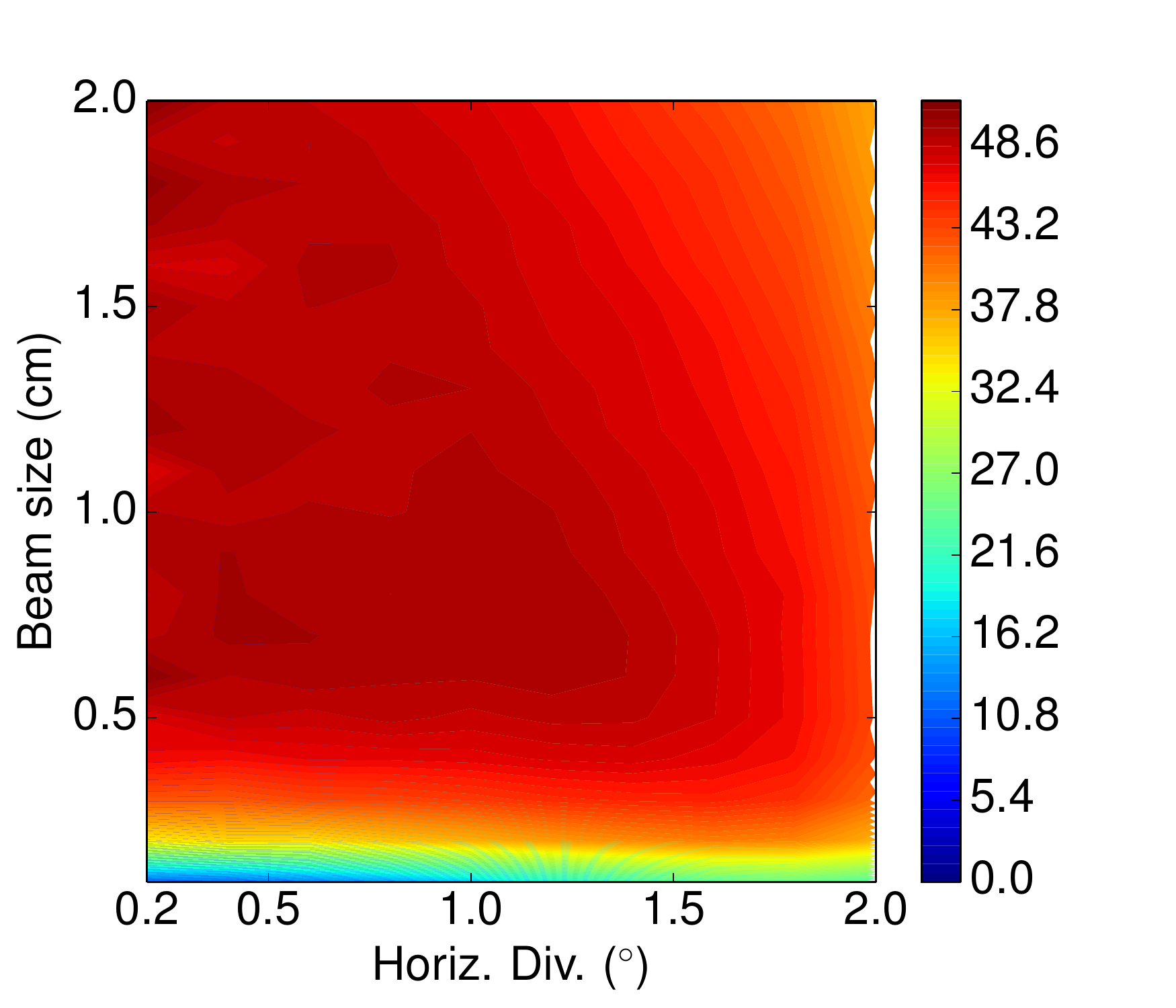}
    \caption{ 
        \label{fig:paramter}
        The BT in dependence of beam size and beam
        divergence for (a) the dual ellipse  and (b) a 2-fold Montel guide for a
        wavelength of $4 \AA$. The ellipse is better for larger sample sizes whereas the
        Montel mirrors are better suited for smaller sample sizes. 
    } 
\end{figure}

\begin{figure}
    \centering
    \includegraphics[width=\linewidth]{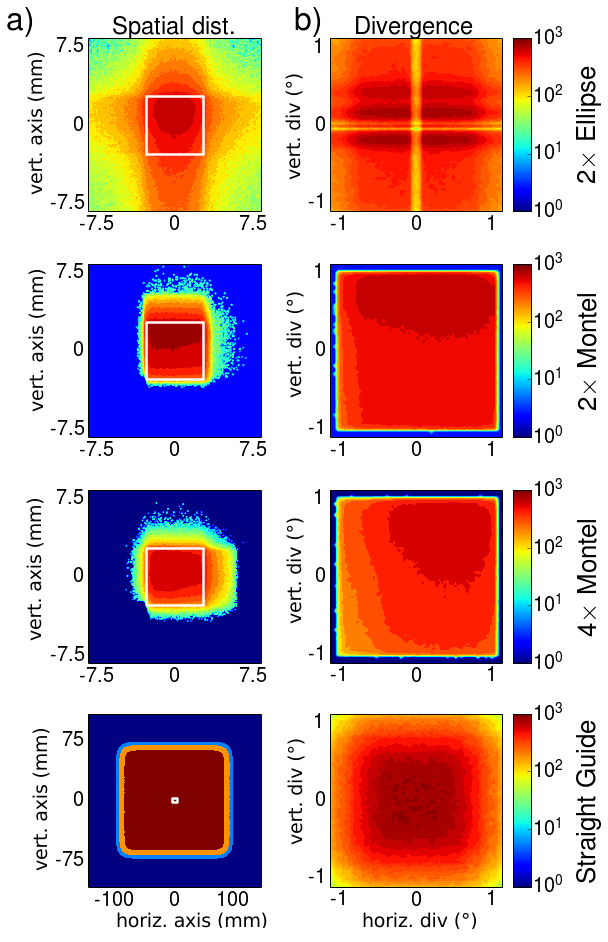}
    \caption{ \label{psd}  Spatial (a) and divergence (b) intensity distribution at the
    sample position for the different guide systems integrated over a flat
    wavelength spectrum of \SI{2}{\AA}~--~\SI{6}{\AA}. The left side (a) shows
    monitor images of the spatial distribution of the neutrons with gravity  and
    the right side (b) the divergence distribution, respectively. The white
    rectangle in (a) marks the sample with a size of $5\times\SI{5}{mm^2}$,
    which corresponds to the opening aperture for the elliptic guide and the
    Montel guide system. The straight guide was simulated without aperture.}
\end{figure}

The major disadvantages of a straight guide are the reflection losses, which increase the shielding effort considerably and the
illumination of the surroundings close to the sample thus increasing the
background and reducing the signal-to-noise ratio (c.f.\ Fig.\ \ref{psd} a and
Tab.~\ref{tab:intensities}).  On the contrary gravity has no effect on the performance of
straight guides (the ratio in Tab.~\ref{tab:intensities} is equal to 1) as it
only leads to slightly larger angles of reflection for neutrons when they are
"hopping" along the guide. As it is relative easy and cheap to produce and
transports large beams over a large wavelength band without big losses it is
widely used in neutron facilities.

The elliptic guide transports nearly all neutrons in a wavelength range of
\SI{2}{\AA}~--~\SI{6}{\AA} (see figure~\ref{fig:paramter}) with a BT up to 72\%,
mainly for larger beam sizes. A detailed analysis of the BT for elliptic guides
has been performed by Klen\o\ et al.\ \cite{Kleno2012}. They obtain a similar
larger BT  for a $\lambda = 1.5$ \AA~and a m=3 guide. As seen from the spatial
distribution in Fig.~{\ref{psd}a} the relatively low SNR of 57\% (see
Tab.~\ref{tab:intensities}) is due to half of the neutrons missing the sample.
Furthermore this type of guides is sensitive to gravity, the beam decreases in
size and in intensity and the integrated intensity on the sample decreases by a
factor of $1.7$.  The elliptic guide is well suited for shorter wavelength for
reducing the background at the sample compared to the straight guide.

With enabled gravity the 2-fold and the 4-fold Montel systems
provide the best performance in the wavelength range of \SI{2}{\AA}~--~\SI{6}{\AA}. They deliver only neutrons which hit the sample giving a very high SNR of 90~\% (see Fig.~\ref{psd}a and Tab.~\ref{tab:intensities}).
For guides composed of Montel mirrors the sensitivity to gravity is less as for elliptic guides.
 As the maximum of the BT shifts
with increasing number of mirrors to longer wavelengths for each application the
appropriate system needs to be chosen.  The 2-fold Montel is ideal
for delivering neutrons with an excellent signal-to-noise ratio to small samples
in a wavelength range between \SI{3}{\AA}~--~\SI{6}{\AA}.

\begin{figure}
    \includegraphics[width=\linewidth]{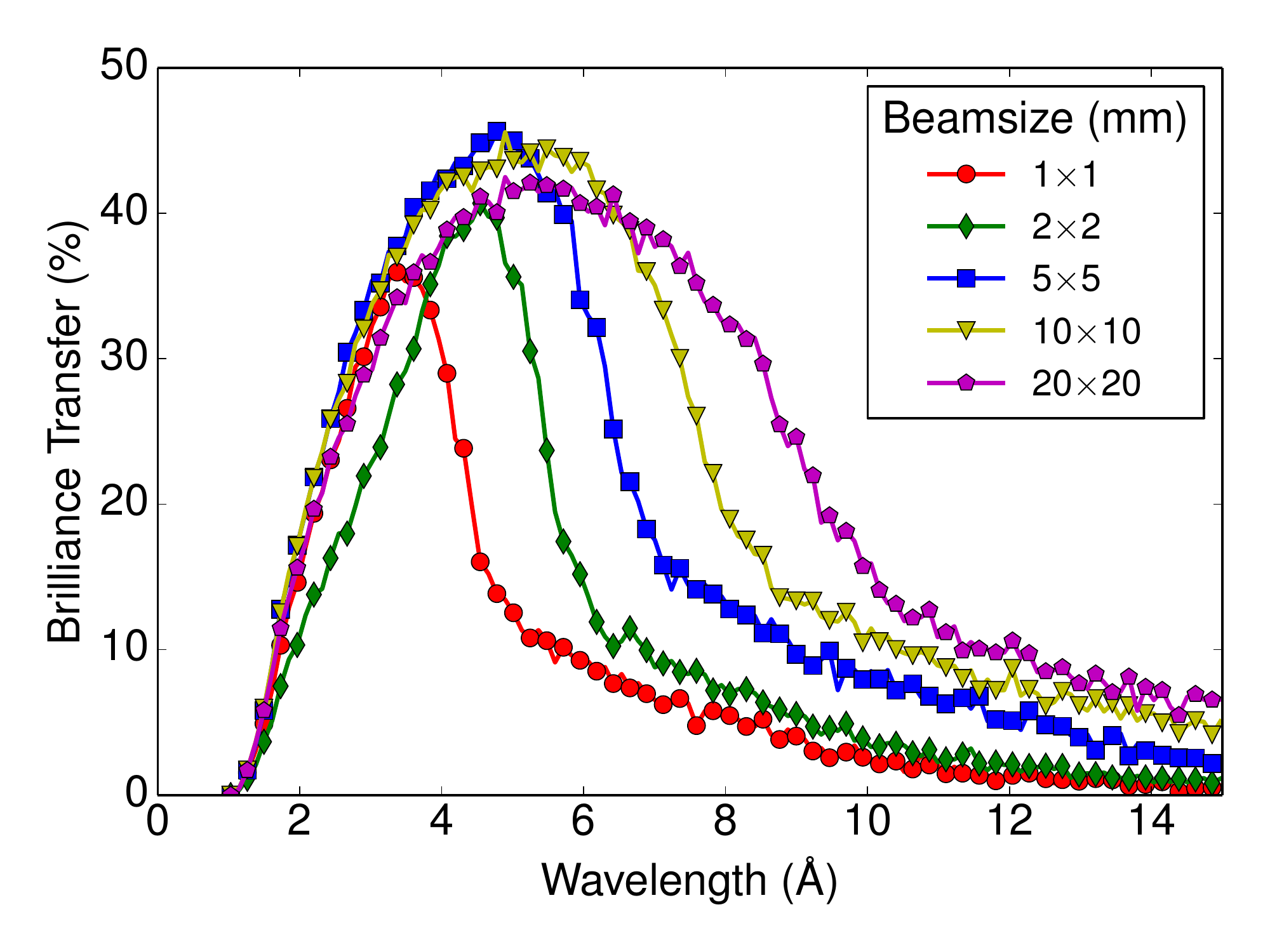}
    \caption{\label{beamsize} Comparison of the BT of a 2-fold Montel system for
    various beam sizes. Effects of gravity are taken into  account. The
aperture at the entry and the detector at the sample position have the same size
as given in the legend. }
\end{figure}

Fig.~\ref{beamsize} shows the performance of the 2-fold Montel system versus
sample size with gravity enabled. The width of the wavelength band
$\Delta \lambda/\lambda$ with high BT increases from $\simeq 0.56$ to $\simeq 1.2$ when the
sample size is increased from 1 mm $\times$ 1 mm to 20 mm $\times$ 20 mm.
The optimum BT is obtained for samples  $5\times\SI{5}{mm^2}$ and \SI{3}{\AA} $\le
\lambda \le$ \SI{6}{\AA}, a parameter range that will be discussed in more
detail in the following, as it will be of great interest for many
instrument designs for the ESS \cite{esstechnologicaldocs}. 

In Fig.~\ref{psd} the spatial and divergence distribution on the sample is given
for a 2-fold Montel mirror with an entry aperture size of $5\times 5\,$mm$^2$ .
The assumed sample size of $5 \times \SI{5}{mm^2}$ is indicated in part a) of
this figure by white rectangles and the maximal intensity  is normalized for
better comparison. Neutrons may not hit the sample i) due to reflection losses
or ii) due to arriving outside the white rectangles thus contributing to the
background.  
 
The top row of the image shows the spatial and divergence distribution for the
dual ellipse system. A considerable part of the neutrons arrive outside the sample. Also note the
gap in the divergence distribution which is caused by the beam blocker.
In rows two and three one can clearly see the effects of an increasing number of mirrors.
For increasing number of Montel
mirrors the homogeneity of the spatial and divergence distribution increases.
The four mirror systems leads to a very homogeneous illumination of the sample area
both for spatial and divergence coordinates. 
The straight guide system shows a very homogenous spatial and divergence
distribution, although the majority of neutrons does not contribute to the
sample illumination.

\section{Conclusions}
We have shown that multiple elliptic guides and Montel mirror systems provide an
efficient neutron transport from the moderator to the sample. As the selection
of the phase space is conducted close to the moderator, the background outside
the biological shielding of the neutron source is massively reduced leading to a
very low background at the sample position. As a side-effect, the costs for
shielding are reduced. Using geometrical optics the brilliance transfer (BT)
can be easily calculated if beam size and gravitational effects are neglected,
which is possible for short guide systems.  Monte-Carlo simulations show that
for the assumed unfavorable conditions we considered, i.e. long flight paths of
\SI{156}{m} and small sample size lead to BTs of 40\%. Almost all of the
neutrons transported through the guide system reach the sample and there will be
practically no background. To quantify this we consider for an example the
thermal beam port H12 at the ILL with its brilliance at $\lambda  = 4~\AA$ of
$\Psi = 4 \times 10^{12} $cm$^{-2} $s$^{-1} \AA^{-1}$ sterad$^{-1}$
\cite{yellowbook1986}.  Using
\begin{equation}
I =  \Psi \times A \times \Omega \times \Delta \lambda \times BT
\end{equation}
and assuming a $\Delta \lambda / \lambda= 2~\AA$ and a horizontal and vertical divergence of $\pm 1^o$, we end up with $2 \times 10^9$ neutrons per second hitting on the sample with an area $A = 5$ mm$^2$. For the ESS similar time averaged intensities with no background can be expected.


For high throughput of large beams  and low background, a
double elliptic guide system using small entrance and exit to reduce the number
of reflections $N > 2$ (i.e. focusing configuration)  may be the optimum choice. 
Finally, when choosing the optimum guide geometry the perfection of the mirrors should be respected.
Presently, Montel mirrors and neutron guides have a waviness of $\simeq
1.0\cdot10^{-5}$ rad and $\simeq 1.0\cdot10^{-4}$ rad, respectively, leading to
a blurring of the beam over a distance of 100 m of 1 mm and 10 mm, respectively.
These values should be compared with the effects of gravity.

In a further study one may also consider more advanced geometries for Montel
optics such as systems being composed of parabolic Montel mirrors at the
entrance and the exit of a guide system connected via a long straight guide
section. Such a geometry may reduce the effects of gravity further.

In the future, it may become possible to build neutron sources based on the
ejection of photo neutrons from halo isomers by means of $\gamma$- and laser
beams, which will provide neutron beams with a very high brilliance
\cite{habs2011} and a small diameter of the order of 0.1 mm. The small beam size leads to short mirrors and therefore effects of gravity become a minor issue.

\section{Acknowledgements}
This work was funded by  the German BMBF under ``Mitwirkung der Zentren der Helmholtz Gemeinschaft und der Technischen Uni\-ver\-sit\"at M\"un\-chen an der Design-Update Phase der ESS, F\"or\-der\-kenn\-zeichen 05E10WO1.''

\bibliography{montel}

\begin{thebibliography}{21}
\providecommand{\natexlab}[1]{#1}
\providecommand{\url}[1]{\texttt{#1}}
\providecommand{\urlprefix}{URL }
\expandafter\ifx\csname urlstyle\endcsname\relax
  \providecommand{\doi}[1]{doi:\discretionary{}{}{}#1}\else
  \providecommand{\doi}[1]{doi:\discretionary{}{}{}\begingroup
  \urlstyle{rm}\url{#1}\endgroup}\fi
\providecommand{\bibinfo}[2]{#2}

\bibitem[{yel(2008)}]{yellowbook1986}
\bibinfo{title}{Yellow Book of the ILL, p~4--6}, \bibinfo{year}{2008}.

\bibitem[{Maier-Leibnitz and Springer(1963)}]{MaierLeibnitz1963217}
\bibinfo{author}{H.~Maier-Leibnitz}, \bibinfo{author}{T.~Springer},
  \bibinfo{title}{The use of neutron optical devices on beam-hole experiments
  on beam-hole experiments}, \bibinfo{journal}{Journal of Nuclear Energy. Parts
  A/B. Reactor Science and Technology}
  \bibinfo{volume}{17}~(\bibinfo{number}{4–5}) (\bibinfo{year}{1963})
  \bibinfo{pages}{217 -- 225}, ISSN \bibinfo{issn}{0368-3230},
  \doi{\bibinfo{doi}{http://dx.doi.org/10.1016/0368-3230(63)90022-3}},
  \urlprefix\url{http://www.sciencedirect.com/science/article/pii/0368323063900223}.

\bibitem[{SwissNeutronics(2014)}]{swissneutronics}
\bibinfo{author}{SwissNeutronics}, \bibinfo{title}{Neutron Supermirrors},
  \urlprefix\url{http://www.swissneutronics.ch/products/coatings.html},
  \bibinfo{year}{2014}.

\bibitem[{Böni et~al.(2010)Böni, Grünauer, and Schanzer}]{boni2010}
\bibinfo{author}{P.~Böni}, \bibinfo{author}{F.~Grünauer},
  \bibinfo{author}{C.~Schanzer}, \bibinfo{title}{Shielding of elliptic guides
  with direct sight to the moderator}, \bibinfo{journal}{Nuclear Instruments
  and Methods in Physics Research Section A: Accelerators, Spectrometers,
  Detectors and Associated Equipment}
  \bibinfo{volume}{624}~(\bibinfo{number}{1}) (\bibinfo{year}{2010})
  \bibinfo{pages}{162 -- 167}, ISSN \bibinfo{issn}{0168-9002},
  \doi{\bibinfo{doi}{http://dx.doi.org/10.1016/j.nima.2010.09.015}},
  \urlprefix\url{http://www.sciencedirect.com/science/article/pii/S0168900210019777}.

\bibitem[{Mezei(1997)}]{Mezei1997}
\bibinfo{author}{F.~Mezei}, \bibinfo{journal}{J. Neutron Res.}
  \bibinfo{volume}{6} (\bibinfo{year}{1997}) \bibinfo{pages}{3}.

\bibitem[{Schanzer et~al.(2004)Schanzer, Böni, Filges, and
  Hils}]{Schanzer200463}
\bibinfo{author}{C.~Schanzer}, \bibinfo{author}{P.~Böni},
  \bibinfo{author}{U.~Filges}, \bibinfo{author}{T.~Hils},
  \bibinfo{title}{Advanced geometries for ballistic neutron guides},
  \bibinfo{journal}{Nuclear Instruments and Methods in Physics Research Section
  A: Accelerators, Spectrometers, Detectors and Associated Equipment}
  \bibinfo{volume}{529}~(\bibinfo{number}{1–3}) (\bibinfo{year}{2004})
  \bibinfo{pages}{63 -- 68}, ISSN \bibinfo{issn}{0168-9002},
  \doi{\bibinfo{doi}{http://dx.doi.org/10.1016/j.nima.2004.04.178}},
  \urlprefix\url{http://www.sciencedirect.com/science/article/pii/S0168900204008605},
  \bibinfo{note}{proceedings of the Joint Meeting of the International
  Conference on Neutron Optics (NOP2004) and the Third International Workshop
  on Position-Sensitive Neutron Detectors (PSND2004)}.

\bibitem[{Cussen et~al.(2013)Cussen, Nekrassov, Zendler, and
  Lieutenant}]{cussen2013}
\bibinfo{author}{L.~D. Cussen}, \bibinfo{author}{D.~Nekrassov},
  \bibinfo{author}{C.~Zendler}, \bibinfo{author}{K.~Lieutenant},
  \bibinfo{title}{Multiple reflections in elliptic neutron guide tubes},
  \bibinfo{journal}{Nuclear Instruments and Methods in Physics Research Section
  A: Accelerators, Spectrometers, Detectors and Associated Equipment}
  \bibinfo{volume}{705}~(\bibinfo{number}{0}) (\bibinfo{year}{2013})
  \bibinfo{pages}{121 -- 131}.

\bibitem[{Ibberson(2009)}]{ibberson2009}
\bibinfo{author}{R.~M. Ibberson}, \bibinfo{title}{{Design and performance of
  the new supermirror guide on HRPD at ISIS}}, \bibinfo{journal}{Nuclear
  Instruments and Methods in Physics Research, Section A: Accelerators,
  Spectrometers, Detectors and Associated Equipment} \bibinfo{volume}{600}
  (\bibinfo{year}{2009}) \bibinfo{pages}{47--49}, ISSN
  \bibinfo{issn}{01689002}, \doi{\bibinfo{doi}{10.1016/j.nima.2008.11.066}}.

\bibitem[{Böni(2014)}]{boeni2014}
\bibinfo{author}{P.~Böni}, \bibinfo{title}{High Intensity Neutron Beams for
  Small Samples}, \bibinfo{journal}{Journal of Physics: Conference Series}
  \bibinfo{volume}{502}~(\bibinfo{number}{1}) (\bibinfo{year}{2014})
  \bibinfo{pages}{012047},
  \urlprefix\url{http://stacks.iop.org/1742-6596/502/i=1/a=012047}.

\bibitem[{Klen{\o} et~al.(2012)Klen{\o}, Lieutenant, Andersen, and
  Lefmann}]{Kleno2012}
\bibinfo{author}{K.~H. Klen{\o}}, \bibinfo{author}{K.~Lieutenant},
  \bibinfo{author}{K.~H. Andersen}, \bibinfo{author}{K.~Lefmann},
  \bibinfo{title}{{Systematic performance study of common neutron guide
  geometries}}, \bibinfo{journal}{Nuclear Instruments and Methods A}
  \bibinfo{volume}{696} (\bibinfo{year}{2012}) \bibinfo{pages}{75--84},
  \urlprefix\url{http://gateway.webofknowledge.com/gateway/Gateway.cgi?GWVersion=2\&SrcAuth=mekentosj\&SrcApp=Papers\&DestLinkType=FullRecord\&DestApp=WOS\&KeyUT=000311570900007
  papers2://publication/doi/10.1016/j.nima.2012.08.027}.

\bibitem[{B\"{o}ni(2008)}]{boni2008}
\bibinfo{author}{P.~B\"{o}ni}, \bibinfo{title}{{New concepts for neutron
  instrumentation}}, \bibinfo{journal}{Nuclear Instruments and Methods in
  Physics Research, Section A: Accelerators, Spectrometers, Detectors and
  Associated Equipment} \bibinfo{volume}{586} (\bibinfo{year}{2008})
  \bibinfo{pages}{1--8}, ISSN \bibinfo{issn}{01689002},
  \doi{\bibinfo{doi}{10.1016/j.nima.2007.11.059}}.

\bibitem[{Zendler et~al.(2014)Zendler, Nekrassov, and Lieutenant}]{zendler2014}
\bibinfo{author}{C.~Zendler}, \bibinfo{author}{D.~Nekrassov},
  \bibinfo{author}{K.~Lieutenant}, \bibinfo{title}{An improved elliptic guide
  concept for a homogeneous neutron beam without direct line of sight},
  \bibinfo{journal}{Nuclear Instruments and Methods in Physics Research Section
  A: Accelerators, Spectrometers, Detectors and Associated Equipment}
  \bibinfo{volume}{746}~(\bibinfo{number}{0}) (\bibinfo{year}{2014})
  \bibinfo{pages}{39 -- 46}, ISSN \bibinfo{issn}{0168-9002},
  \doi{\bibinfo{doi}{http://dx.doi.org/10.1016/j.nima.2014.01.044}},
  \urlprefix\url{http://www.sciencedirect.com/science/article/pii/S0168900214000953}.

\bibitem[{Klen{\o} et~al.(2011)Klen{\o}, Willendrup, Knudsen, and
  Lefmann}]{kleno2011}
\bibinfo{author}{K.~H. Klen{\o}}, \bibinfo{author}{P.~K. Willendrup},
  \bibinfo{author}{E.~Knudsen}, \bibinfo{author}{K.~Lefmann},
  \bibinfo{title}{{Eliminating line of sight in elliptic guides using
  gravitational curving}}, \bibinfo{journal}{Nuclear Instruments and Methods A}
  \bibinfo{volume}{634}~(\bibinfo{number}{1}) (\bibinfo{year}{2011})
  \bibinfo{pages}{0},
  \urlprefix\url{http://pubget.com/site/paper/pgtmp\_1ba3576f617a94c74d2d1c252ff4f5df?institution=
  papers2://publication/doi/10.1016/j.nima.2010.06.261}.

\bibitem[{Montel(1957)}]{Montel:1957}
\bibinfo{author}{M.~Montel}, \bibinfo{title}{X-ray microscopy with catamegonic
  roof mirrors, X-ray microscopy and microradiography},
  \bibinfo{publisher}{Academic Press}, \bibinfo{year}{1957}.

\bibitem[{Ice et~al.(2009)Ice, Pang, Tulk, Molaison, Choi, Vaughn, Lytle,
  Takacs, Andersen, Bigault, and Khounsary}]{Ice:he5452}
\bibinfo{author}{G.~E. Ice}, \bibinfo{author}{J.~W.~L. Pang},
  \bibinfo{author}{C.~Tulk}, \bibinfo{author}{J.~Molaison},
  \bibinfo{author}{J.-Y. Choi}, \bibinfo{author}{C.~Vaughn},
  \bibinfo{author}{L.~Lytle}, \bibinfo{author}{P.~Z. Takacs},
  \bibinfo{author}{K.~H. Andersen}, \bibinfo{author}{T.~Bigault},
  \bibinfo{author}{A.~Khounsary}, \bibinfo{title}{Design challenges and
  performance of nested neutron mirrors for microfocusing on SNAP},
  \bibinfo{journal}{Journal of Applied Crystallography}
  \bibinfo{volume}{42}~(\bibinfo{number}{6}) (\bibinfo{year}{2009})
  \bibinfo{pages}{1004--1008}, \doi{\bibinfo{doi}{10.1107/S0021889809037595}},
  \urlprefix\url{http://dx.doi.org/10.1107/S0021889809037595}.

\bibitem[{Stahn et~al.(2012)Stahn, Filges, and Panzner}]{stahn2012}
\bibinfo{author}{J.~Stahn}, \bibinfo{author}{U.~Filges},
  \bibinfo{author}{T.~Panzner}, \bibinfo{title}{Focusing specular neutron
  reflectometry for small samples}, \bibinfo{journal}{The European Physical
  Journal Applied Physics} \bibinfo{volume}{58}, ISSN
  \bibinfo{issn}{1286-0050}, \doi{\bibinfo{doi}{10.1051/epjap/2012110295}}.

\bibitem[{Weber et~al.(2013)Weber, Brandl, Georgii, Häußler, Weichselbaumer,
  and Böni}]{Weber201371}
\bibinfo{author}{T.~Weber}, \bibinfo{author}{G.~Brandl},
  \bibinfo{author}{R.~Georgii}, \bibinfo{author}{W.~Häußler},
  \bibinfo{author}{S.~Weichselbaumer}, \bibinfo{author}{P.~Böni},
  \bibinfo{title}{Monte-Carlo simulations for the optimisation of a TOF-MIEZE
  instrument}, \bibinfo{journal}{Nuclear Instruments and Methods in Physics
  Research Section A: Accelerators, Spectrometers, Detectors and Associated
  Equipment} \bibinfo{volume}{713}~(\bibinfo{number}{0}) (\bibinfo{year}{2013})
  \bibinfo{pages}{71 -- 75}, ISSN \bibinfo{issn}{0168-9002},
  \doi{\bibinfo{doi}{http://dx.doi.org/10.1016/j.nima.2013.03.010}}.

\bibitem[{Stahn et~al.(2011)Stahn, Panzner, Filges, Marcelot, and
  Böni}]{stahn2011}
\bibinfo{author}{J.~Stahn}, \bibinfo{author}{T.~Panzner},
  \bibinfo{author}{U.~Filges}, \bibinfo{author}{C.~Marcelot},
  \bibinfo{author}{P.~Böni}, \bibinfo{title}{Study on a focusing,
  low-background neutron delivery system}, \bibinfo{journal}{Nuclear
  Instruments and Methods in Physics Research Section A: Accelerators,
  Spectrometers, Detectors and Associated Equipment}
  \bibinfo{volume}{634}~(\bibinfo{number}{1, Supplement})
  (\bibinfo{year}{2011}) \bibinfo{pages}{S12 -- S16}, ISSN
  \bibinfo{issn}{0168-9002},
  \doi{\bibinfo{doi}{http://dx.doi.org/10.1016/j.nima.2010.06.221}},
  \urlprefix\url{http://www.sciencedirect.com/science/article/pii/S0168900210014117},
  \bibinfo{note}{proceedings of the International Workshop on Neutron Optics
  {NOP2010}}.

\bibitem[{Willendrup et~al.(2014)Willendrup, Knudsen, Lefmann, Farhi, and
  Filges}]{mcstashomepage}
\bibinfo{author}{P.~Willendrup}, \bibinfo{author}{E.~Knudsen},
  \bibinfo{author}{K.~Lefmann}, \bibinfo{author}{E.~Farhi},
  \bibinfo{author}{U.~Filges}, \bibinfo{title}{McStas -- A neutron ray-trace
  simulation package}, \urlprefix\url{http://mcstas.org}, \bibinfo{year}{2014}.

\bibitem[{ESS(2013)}]{esstechnologicaldocs}
\bibinfo{author}{ESS}, \bibinfo{title}{Scientific \& Technological
  Documentation},
  \urlprefix\url{http://europeanspallationsource.se/scientific-technological-documentation},
  \bibinfo{year}{2013}.

\bibitem[{Habs et~al.(2011)Habs, Gross, Thirolf, and Böni}]{habs2011}
\bibinfo{author}{D.~Habs}, \bibinfo{author}{M.~Gross},
  \bibinfo{author}{P.~Thirolf}, \bibinfo{author}{P.~Böni},
  \bibinfo{title}{Neutron halo isomers in stable nuclei and their possible
  application for the production of low energy, pulsed, polarized neutron beams
  of high intensity and high brilliance}, \bibinfo{journal}{Applied Physics B}
  \bibinfo{volume}{103}~(\bibinfo{number}{2}) (\bibinfo{year}{2011})
  \bibinfo{pages}{485--499}, ISSN \bibinfo{issn}{0946-2171},
  \doi{\bibinfo{doi}{10.1007/s00340-010-4276-3}},
  \urlprefix\url{http://dx.doi.org/10.1007/s00340-010-4276-3}.

\end{thebibliography}
\end{document}